\documentclass[aps,pra,superscriptaddress,twocolumn,floatfix,a4paper]{revtex4}

\usepackage{graphicx,graphics,epsfig}   
\usepackage{dcolumn}    
\usepackage{bm}         
\usepackage{amsmath}    
\usepackage{verbatim}   
\usepackage{color}      
\usepackage{subfigure}  
\usepackage{times,natbib}
\usepackage{amsmath,amsfonts,amssymb,graphics,graphics,color,times}

\usepackage{latexsym}
\usepackage{amsmath}
\usepackage{amssymb}
\usepackage{amsfonts}
\usepackage{amsthm}
\usepackage{mathrsfs}
\usepackage{color,verbatim,graphics}
\usepackage{psfrag}
\DeclareMathAlphabet{\mathrsfs}{U}{rsfs}{m}{n}
\DeclareMathAlphabet{\mathpzc}{OT1}{pzc}{m}{it}
\DeclareMathAlphabet{\matheus}{U}{eus}{m}{n}
\DeclareMathAlphabet{\mathbbold}{U}{bbold}{m}{n}

\newcommand{\ba}{\begin{eqnarray}}
\newcommand{\ea}{\end{eqnarray}}
\newcommand{\ban}{\begin{eqnarray*}}
\newcommand{\ean}{\end{eqnarray*}}
\newcommand{\Tr}{\operatorname{Tr}}

\begin{document}

\title{Certifying nonlocality from separable marginals}

\author{Tam\'as V\'ertesi}
\affiliation{Institute for Nuclear Research, Hungarian Academy of
Sciences, H-4001 Debrecen, P.O. Box 51, Hungary}

\author{Wies\l aw~Laskowski}
\affiliation{Institute of Theoretical Physics and Astrophysics,
University of Gda\'nsk, 80-952 Gda\'nsk, Poland}

\author{K\'aroly F. P\'al}
\affiliation{Institute for Nuclear Research, Hungarian
Academy of Sciences, H-4001 Debrecen, P.O. Box 51, Hungary}



\begin{abstract}
Imagine three parties, Alice, Bob, and Charlie, who share a state
of three qubits such that all two-party reduced states $A$-$B$, $A$-$C$,
and $B$-$C$ are separable. Suppose that they have information only about
these marginals but not about the global state. According to
recent results, there exists an example for a set of three separable
two-party reduced states that is only compatible with an
entangled global state. In this paper, we show a stronger result,
by exhibiting separable two-party reduced states $A$-$B$, $A$-$C$, and
$B$-$C$, such that any global state compatible with these marginals is
nonlocal. Hence, we obtain that nonlocality of multipartite states can be 
certified from information only about separable marginals.
\end{abstract}

\maketitle

\section{Introduction}\label{intro}

Entanglement \cite{horo} and nonlocality \cite{nonloc} are two
defining aspects of quantum mechanics providing powerful resources
for numerous applications in quantum information science.
Although long ago they were thought to be two facets of the
same phenomenon, these two notions of inseparability have turned
out to be quite different \cite{diff}. Crucially, due to Bell's
theorem \cite{bell}, distant parties sharing an entangled quantum
state can generate nonlocal correlations, witnessed by violation
of a Bell inequality, which rules out any local realistic model.
However, it is difficult to fully identify the set of entangled
states which are nonlocal, i.e., give rise to Bell violation.

In the simplest bipartite scenario, for instance, there exists a
family of quantum states, the so-called Werner states
\cite{local2}, which are entangled but nevertheless are local (i.e.,
admit a local realistic model for any single-shot measurement).
Similarly, in the tripartite case, there exists a family of
entangled three-qubit states having a local realistic model for
any single-shot von Neumann measurement \cite{local3}.
Interestingly, some of these three-qubit states are genuinely
multipartite entangled, representing a very strong form of
multipartite entanglement. Conversely, it has been recently shown
that a three-qubit bound entangled state (where entanglement in
the system presents in a very weak, almost invisible form)
exhibits nonlocality; that is, it violates a tripartite Bell
inequality \cite{VB}.

This selection of works already suggests that the relation between
entanglement and nonlocality is very subtle. In our present work,
we wish to give a further example linking the two concepts to each
other in an intriguing way. The question we pose is the following.
Does there exist a three-party system with a set of two-party
separable reduced states for which any global state compatible
with these reduced states is nonlocal? Note that two related
questions have already been addressed: (i) Can one deduce that a
global state is entangled from the observation of separable
reduced states~\cite{sep1,sep2}? (ii) Can one deduce that a
global state is non-local from the observation of local marginal
correlations~\cite{sep2,LMPW}?

To both questions the answer turns out to be yes. However, as
posed above, our goal in this paper is to answer a question which
is strictly stronger than both questions above. In particular, we
wonder if there exist reduced states which are nonentangled where,
however, any three-qubit state compatible with these marginals is
nonlocal. In this case, subcorrelation Bell inequalities
\cite{BSV,LMPW,sep2,detect} come to our aid. These types of Bell
inequalities do not involve full-correlation terms (that is,
correlation terms which consist of all parties), and in the special
case of three parties they contain only two- and one-body
expectation values.

As a starting point for our study, we exhibit in
section~\ref{sepstate} a tripartite quantum state which has
separable two-party reduced states. Then, we introduce in
section~\ref{two-body} a subcorrelation Bell inequality (involving
only one- and two-body mean values) which is violated by the
above quantum state provided well-chosen measurements are
performed on it. Note that due to the special form of our Bell
inequality, the quantum expectation values, and, consequently the
quantum violation of the Bell inequality depend only on the
two-party reduced states and not on the global state itself. Then,
we obtain that any extension of the above set of separable
two-party reduced states to a global state results in a nonlocal
global state. This already implies our main result stated in the
abstract. However, it turns out that the Bell violation with the
above two-party reduced states is very small (in the range of
$10^{-2}$) and therefore very sensitive to noise which
arises inevitably in any experimental setup.

In order to propose a scheme which is more robust to noise, we
give a simple method in section~\ref{compat} based on semidefinite
programming (SDP), which allows us to decide whether a global state
is fully determined by its reduced states. By applying this method
to our set of three two-party reduced states introduced in
section~\ref{sepstate}, we find out that these marginals in fact
fully determine the global three-party state. This implies that
the violation of an arbitrary three-party Bell inequality
(possibly consisting of all-correlation terms as well) signals the
nonlocality of any global state compatible with the two-party
reductions of the global state. Therefore, in the following we do
not have to restrict ourselves to the study of two-body Bell
inequalities. Indeed, we provide in section~\ref{three-body} a
three-party Bell inequality which is violated by a large amount
using our unique state with separable marginals. The relatively
big violation suggests that our example is promising from the
viewpoint of possible experimental implementation as well.

\section{A family of three-qubit states}\label{sepstate}

Our starting point is the following family of states:
\begin{align}
\label{rhoabc}
\varrho=&p_0|0\rangle\langle 0|\otimes|\psi_0\rangle\langle\psi_0|\nonumber\\
&+|1\rangle\langle
1|\otimes(p_1|\psi_1\rangle\langle\psi_1|+p_2|\psi_2\rangle\langle\psi_2|),
\end{align}
where Alice holds the first qubit and the pure two-qubit states
$|\psi_i\rangle$, $i=0,1,2$, possessed by Bob and Charlie have the special
parametric form
\begin{align}
\label{psii}
|\psi_0\rangle =& \cos\alpha|00\rangle + \sin\alpha|11\rangle,\nonumber\\
|\psi_1\rangle =& (\cos\beta|0\rangle + \sin\beta|1\rangle)\otimes (\cos\gamma|0\rangle + \sin\gamma|1\rangle),\nonumber\\
|\psi_2\rangle =& \frac{1}{\sqrt 2}(\sin\delta|00\rangle +
\cos\delta|01\rangle + \cos\delta|10\rangle
-\sin\delta|11\rangle).
\end{align}
Note that $|\psi_1\rangle$ is a product state, whereas
$|\psi_0\rangle$ is a partially entangled state for generic angle
$\alpha$ and $|\psi_2\rangle$ is a maximally entangled state. Also
note that, due to construction, the state is biseparable with
respect to cut $A|BC$, which implies that both $\varrho_{AB}$ and
$\varrho_{AC}$ two-party reduced states are separable (for a
review of different notions of separability, we refer the reader
to Ref.~\cite{GT}). On the other hand, tracing out
Alice's qubit we get the reduced state
$\rho_{BC}=\sum_{i=0,1,2}p_i|\psi_i\rangle\langle\psi_i|$. Let us
now fix weights $p_i$,
\begin{align}
\label{pi}
p_0 &= 0.759101,\nonumber\\
p_1 &= 0.015596,\nonumber\\
p_2 &= 0.225303,
\end{align}
and the angles
\begin{align}
\label{angle}
\alpha &= 0.093586,\nonumber\\
\beta &= 1.228106,\nonumber\\
\gamma &= 1.063034,\nonumber\\
\delta &= 0.050725.
\end{align}
entering the three-party state~(\ref{rhoabc}) along with two-qubit pure states~(\ref{psii}) held by Bob and Charlie.

In the following, let us denote by $\varrho^*$ the
state~(\ref{rhoabc}) with the specially chosen parameters
(\ref{psii}),(\ref{pi}), and (\ref{angle}). Using the Peres transposition
map \cite{peres}, we find that the two-party reduced state
$\rho_{BC}=\sum_{i=0,1,2}p_i|\psi_i\rangle\langle\psi_i|$ of the global state $\varrho^*$ is separable as
well. Hence, we can conclude that all three two-party reduced
states of the state $\varrho^*$ are separable.

\section{Two-body Bell inequality}\label{two-body}

We now present a three-party Bell inequality, where each party has
a maximum of two possible binary measurements $A_i, B_i, C_i$, $i=1,2$.
The Bell expression consists of only single party marginals and
two-body correlation terms defined by the following sum of
expectation values:
\begin{equation}
\label{BI} BI = -A_1 + (B_1 - B_2 - C_2)(1+A_1) + CHSH_{BC} \le 3,
\end{equation}
where the last term on the left-hand side defines the Clauser-Horne-Shimony-Holt (CHSH)
quantity~\cite{CHSH},
\begin{equation}
\label{chsh}
CHSH_{BC} = B_1C_1 + B_1C_2 + B_2C_1 - B_2C_2.
\end{equation}
Let us briefly mention that the above Bell inequality~(\ref{BI})
defines a facet of the polytope of classical correlations which
now lives in the reduced space of single- and two-party correlators
(i.e., neglecting correlators of order 3). One may arrive at
the above Bell inequality, for instance, by means of a geometric
approach similar to the one used in \cite{sep2}.

Let us remark that Alice in the above Bell inequality~(\ref{BI})
performs only a single measurement $A_1$. In the classical case,
(that is, in case of local realistic models) the Bell
expression~(\ref{BI}) is bounded by the value of 3. However, by
performing suitable measurements on the state $\varrho^*$, it
becomes possible to beat this bound. Here we show it by giving the
actual measurements. All of them are of equatorial von
Neumann type, which can be written in the form $A_i =
\cos\theta^a_i \sigma_z + \sin\theta^a_i \sigma_x$, where
$\sigma_x$ and $\sigma_z$ are Pauli matrices. The measurements
$B_i$, $C_i$ for Bob and Charlie are denoted analogously. The
corresponding measurement angles are defined by
\begin{align}
\label{thetas}
\theta^a_1 &= 0,\nonumber\\
\theta^b_1 &= 0.320997,\nonumber\\
\theta^c_1 &= 1.442524,\nonumber\\
\theta^b_2 &= 2.707329,\nonumber\\
\theta^c_2 &=-3.108820.
\end{align}
Indeed, the  measurements defined by the angles~(\ref{thetas})
acting on the state $\varrho^*$ lead to the value of $Q=3.017583$ 
in the Bell expression~(\ref{BI}), giving rise to a small (but
nonzero) violation of the inequality.

In order to arrive at the above Bell violation, we applied the
simplex uphill method~\cite{nm} to find the best measurement
operators and the state with the given form~(\ref{rhoabc})
fulfilling the condition that the two-qubit marginal $\rho_{BC}$ is
separable. This latter condition was imposed by the simple two-qubit
separability condition \cite{adh}, requiring that a two-qubit state
$\rho_{BC}$ is separable if and only if $\det(\rho_{BC}^{T_B})\ge
0$, where the operation $T_B$ denotes partial
transposition~\cite{peres}.

Note that in the case of optimality, the value of the angle $\delta$
in Eq.~(\ref{angle}) is close to zero; hence, the maximally
entangled state $|\psi_2\rangle$ in (\ref{psii}) is close to the
Bell state $|\Psi^+\rangle=\left(|01\rangle + |10\rangle\right)/\sqrt 2$. 
By fixing the form of state $|\psi_2\rangle$ to be state $|\Psi^+\rangle$, 
one gets a Bell violation of
$3.017454$, 
which is only slightly lower than the optimal one presented above
with the parametric form of $|\psi_2\rangle$.

On the other hand, one may wonder what the largest quantum
violation is if one does not stick to the form of the family of
states~(\ref{rhoabc}) but one allows the most general form of a
three-qubit state with separable two-qubit marginals. In that
case, using a seesaw-type iteration technique \cite{seesaw}, the
best state found gives the slightly higher quantum violation of
$3.017924$. 

Let us stress again the unusual feature of the Bell
inequality~(\ref{BI}), namely, that Alice performs only one
measurement on her share of the quantum state. In fact, this
measurement acts as a filter, heralding the desired entangled
state for the remaining two parties. Let us next analyze the Bell
violation from this perspective by giving an alternative way
to arrive at the quantum value of $Q=3.017583$ 
obtained above with the state $\varrho^*$ and particular measurement
angles~(\ref{thetas}).

Alice, by measuring in the standard basis, which corresponds to
the observable $A_1=\sigma_z$, will collapse $\varrho^*$ into
another state. In particular, whenever the result is $A_1=+1$,
whichoccurs with a probability of $p_0$, the projected state becomes
\begin{equation}
\label{rho1} \varrho^+_{BC}=|\psi_0\rangle\langle\psi_0|,
\end{equation}
whereas for the outcome $A_1=-1$, which occurs with a probability of
$1-p_0$, the projected state becomes
\begin{equation}
\label{rho2}
\varrho^-_{BC}=\frac{p_1|\psi_1\rangle\langle\psi_1|+p_2|\psi_2\rangle\langle\psi_2|}{p_1+p_2},
\end{equation}
where we have written both states in a normalized form. Similarly,
the three-party Bell inequality~(\ref{BI}) traces back to two
different two-party Bell inequalities depending on the outcomes
$A_1=\pm 1$,
\begin{align}
BI_+ &= BI(A_1 = +1) \nonumber\\
&= 2(B_1-B_2-C_2) -1 + CHSH_{BC}\le 3,
\nonumber\\
BI_- &= BI(A_1 = -1) = CHSH_{BC} +1 \le 3,
\end{align}
where $CHSH_{BC}$ is defined by Eq.~(\ref{chsh}). Above, the $BI_{\pm}$ 
expressions are obtained by substituting $A_1=\pm 1$ into the Bell expression~(\ref{BI}). 
Our task now is to compute the overall quantum Bell value $Q$ by weighting the
probability of occurrences of the two distinct cases, $Q_+ =
\Tr{(\rho^+_{BC} BI_+)}=2.898134$ 
and $Q_- = \Tr{(\rho^-_{BC} BI_-)}= 3.393981$, 
\begin{equation}
Q = p_0 Q_+ + (1-p_0) Q_- = 3.017583  > 3. 
\end{equation}
Despite the fact that only the second inequality $BI_-$ is
violated, due to the non-negligible probability $(1-p_0)=
0.240899$ 
of the $A_1=-1$ outcome occurring, we get the net violation of $Q
=3.017583$ 
reported above.

Let us next analyze how economic the above devised Bell test is
regarding the number of settings and the state used. First let us
look at the number of settings in Eq.~(\ref{BI}). Alice has one
setting, whereas the other two parties can choose between two
alternative settings. By removing one setting from any of the
parties we get either a trivial or, effectively, a two-party Bell
inequality. In neither case can we arrive at the conclusion that
separable two-party marginals imply nonlocal quantum
correlations. So, regarding the number of settings, the Bell
inequality~(\ref{BI}) defines a minimal construction.

Regarding the state, suppose that we set to zero the small $p_1$ weight
defined by~(\ref{pi}) in the state $\varrho^*$. Then,
there will be at most two terms in the eigendecomposition of the
two-party marginal
$\varrho_{BC}=\sum_{i=0,1,2}p_i|\phi_i\rangle\langle\phi_i|$.
Then, it is known that for any natural measure the set of $(2\times
2)$-dimensional separable states occupies a nonzero
volume~\cite{ZHSL}. However, due to a recent work \cite{RW}, the
respective volume is zero for rank-2 states, such as in the case of
the above $\varrho_{BC}$ with $p_1=0$. This implies that for
$p_1=0$ the reduced state $\varrho_{BC}$ almost certainty becomes entangled.
Hence, we found that the small nonzero component
$p_1$ takes care of the separability of the reduced state
$\varrho_{BC}$ of $\varrho^*$. This means that the rank-3
biseparable state~$\varrho^*$ with nonzero weight $p_1$ is also a
minimal construction in terms of the number of pure-state
decompositions. However, regarding the global state, we assumed the
special form of~(\ref{rhoabc}), and it remains an open question
whether rank-2 (or even rank-1) genuinely tripartite entangled states \cite{GT}
would suffice to prove our result.

As stated before, the example analyzed in this section already
implies the existence of non-entangled two-party reduced states
which are only compatible with non-local global states. However,
by looking more closely at the state $\varrho^*$, it will turn out
that its three separable bipartite reduced states define a unique
extension to a global state (which is the state $\varrho^*$
itself). We find this result via a compatibility test which will
be described in section~\ref{compat}. Then, based on the
uniqueness property of the global three-qubit state $\varrho^*$,
violation of a generic three-party Bell inequality with the state
$\varrho^*$ demonstrates the existence of separable two-party
reduced states that are compatible with nonlocal global states.
Indeed, in section ~\ref{three-body} we find a large violation of a
three-party Bell inequality with the particular state $\varrho^*$,
which implies our stated result.

\section{Compatibility test of the state}\label{compat}

In this section, we show a simple method based on SDP which allows us 
to decide whether a given three-qubit state is fully determined by its 
two-qubit reduced states. The method below is related in spirit to 
the SDP used in Refs.~\cite{hall,sep2} and can be easily generalized to 
higher-dimensional states and more particles as well. 
However, we conjecture that the complexity of the problem will increase
rapidly with the dimension of the state and the number of parties involved.

First, note that any three-qubit density matrix $\rho$ can be
expressed in a tensor form,
\begin{equation}
\rho=\frac{1}{8}\sum_{i_1,i_2,i_3=0}^{3}T_{{i_1},{i_2},{i_3}}\sigma_{i_1}\otimes\sigma_{i_2}\otimes
\sigma_{i_3}, \label{cortensor}
\end{equation}
where $\sigma_{i_k} \in \{\openone, \sigma_1, \sigma_2, \sigma_3
\}$ are the Pauli matrices of the $k$th observer. On the other
hand, the tensor components of a three-qubit state can be readily
obtained by the expectation values,
\begin{equation}
T_{{i_1},{i_2},{i_3}}= \Tr{\rho
\sigma_{i_1}\otimes\sigma_{i_2}\otimes \sigma_{i_3}}.
\label{tcorr}
\end{equation}

In particular, let us denote by $T^*_{i_1,i_2,i_3}$ the tensor
components of our particular state $\varrho^*$ defined by Eqs.~(\ref{rhoabc}),
(\ref{psii}),(\ref{pi}), and (\ref{angle}). Also, note that
the $B$-$C$ two-party reduced state of a general three-qubit state
$\rho$ can be expressed as
\begin{equation}
\Tr_A{\rho}=\frac{1}{4}\sum_{i_2,i_3=0}^{3}T_{{0},{i_2},{i_3}}\sigma_{i_2}\otimes\sigma_{i_3},
\label{cortensorbc}
\end{equation}
and analogous expressions hold for the other bipartite states $A$-$B$
and $B$-$C$ as well. Let us now solve separately the following two SDP
problems for all $i_1,i_2,i_3 = 1,2,3$:

\begin{equation}
\label{primalmax}
\begin{aligned}
T^U_{i_1,i_2,i_3} =  &\; {\text{maximize}} & & T_{i_1,i_2,i_3} \\
&\ \ \ \ \rho, T&&\\
& \text{subject to} & & \rho \succeq 0,\\
& && T_{0,j_2,j_3} = T^*_{0,j_2,j_3}\\
& && T_{j_1,0,j_3} = T^*_{j_1,0,j_3}\\
& && T_{0,j_2,j_3} = T^*_{0,j_2,j_3} \\
& && \forall j_1,j_2,j_3=0,1,2,3\\
\end{aligned}
\end{equation}
and
\begin{equation}
\label{primalmin}
\begin{aligned}
T^L_{i_1,i_2,i_3} =  &\; {\text{minimize}} & & T_{i_1,i_2,i_3} \\
&\ \ \ \ \rho, T&&\\
& \text{subject to} & & \rho \succeq 0,\\
& && T_{0,j_2,j_3} = T^*_{0,j_2,j_3}\\
& && T_{j_1,0,j_3} = T^*_{j_1,0,j_3}\\
& && T_{0,j_2,j_3} = T^*_{0,j_2,j_3} \\
& && \forall j_1,j_2,j_3=0,1,2,3\\
\end{aligned}
\end{equation}

The above SDP optimization problems (note that $\rho$ is a
linear function of the tensor components $T_{i_1,i_2,i_3}$) are
actually the same problems with the only difference being that in the
first case a maximization is carried and in the second case a minimization is
carried out.

Having solved the SDP problem with the particular $T^*$ components
coming from the state $\varrho^*$ and making use of
formula~(\ref{cortensorbc}), we find that
$T^L_{i_1,i_2,i_3}=T^U_{i_1,i_2,i_3}$ for all $i_1,i_2,i_3=1,2,3$
up to a precision of $\sim10^{-10}$, which is roughly the numerical
accuracy of our SDP solver SeDuMi \cite{sturm}. Hence, we conclude
that state~$\varrho^*$ is completely determined by its
two-party reduced states up to high numerical precision. In other
words, all the information of state~$\varrho^*$ is stored
within its two-party reduced states. It is interesting to note
that states with such a property are generic among multipartite pure
states \cite{LPW,LW}. In particular, it was shown by Jones and
Linden \cite{JL} that generic $N$-party pure quantum states (with
equidimensional subsystems) are uniquely determined by the
reduced states of just over half the parties. For a special set of
multipartite states, the so-called $n$-qubit ring cluster states,
an even stronger result have been obtained by T\'oth et al.
\cite{sep1}, who proved that for $n\ge 6$ all neighboring three-party
reduced states are separable and determine uniquely the global
state. However, we are not aware of such results from the
literature in the case of mixed three-qubit states.

\section{Generic three-party Bell inequalities}\label{three-body}

In the previous section, we have seen that the two-party marginals
of the state~$\varrho^*$ defined by Eqs.~(\ref{rhoabc}), (\ref{psii}), 
(\ref{pi}), and (\ref{angle}) determine the
state completely; hence it is legitimate to use generic Bell
inequalities to test the nonlocal nature of the state~$\varrho^*$.
Namely, if we find a violation of a three-party Bell inequality
(possibly consisting of three-body terms as well) with our
state~$\varrho^*$, we can be certain that the only global state
compatible with the two-party marginals of state~$\varrho^*$
is nonlocal. Our goal now is to find a three-party Bell inequality
which gives the biggest $Q/L$ ratio, where $L$ defines the local
bound on the Bell inequality in question and $Q$ is the maximum
quantum value attainable by using state~$\varrho^*$. The
magnitude of the $Q/L$ ratio indicates how useful the Bell
inequality is for our purposes.

Let us first pick the Mermin inequality \cite{mermin}, which
consists of three-party correlation terms,
\begin{equation}
\label{Mermin} M = -A_1B_1C_1 + A_1B_2C_2 + A_2B_1C_2 + A_2B_2C_1
\le 2.
\end{equation}
This is equivalent to number 2 in the complete list of two-setting
three-party Bell inequalities collected by Sliwa \cite{Sliwa}. Let
us now choose the following settings,
\begin{align}
\label{merminsettings}
&A_1 = A_2 = \sigma_z,\nonumber\\
&B_1 = \sigma_z, B_2 = \sigma_y,\nonumber\\
&C_1 = \cos\theta_1\sigma_z + \sin\theta_1\cos\theta_2\sigma_x + \sin\theta_1\sin\theta_2\sigma_y,\nonumber\\
&C_2 = -\cos\theta_1\sigma_z - \sin\theta_1\cos\theta_2\sigma_x +
\sin\theta_1\sin\theta_2\sigma_y.
\end{align}
with $\theta_1=3.500760$ and $\theta_2=1.605042$.
Note the optimal settings are not on the $XY$ plane as usual for the
$GHZ=(1/\sqrt 2)(|000\rangle+|111\rangle)$ state \cite{GHZ}. With
our settings~(\ref{merminsettings}), we get a quantum violation
of $Q = 2.086929$.
However, the optimal quantum violation with the same state
$\varrho^*$ but allowing completely general settings is marginally
bigger, given by $Q = 2.087190$.

By listing all the inequalities in Sliwa's set, number 4 happens to
give the biggest $Q/L$ ratio. The inequality looks as follows:
\begin{equation}
\label{S4} S_4 = (1-A_1)CHSH_{BC}+2A_1 \le 2,
\end{equation}
where $CHSH_{BC}$ is defined by Eq.~(\ref{chsh}). 
Compared to inequality~(\ref{BI}), Alice still performs a single measurement, 
but the inequality now contains three-body terms as well.
The quantum maximum of (\ref{S4}) using state~$\varrho^*$
turns out to be $Q=2.334184$,
which easily follows from our previous analysis in section~\ref{two-body}.

Namely, let Alice measure in the standard basis. With probability
$p_0=0.759101$ 
she projects the state on $|\psi_0\rangle$ [defined
by~(\ref{rho1})] and with probability $1-p_0$ on state
(\ref{rho2}). Then, if Alice gets outcome $A_1=+1$, we obtain the
trivial two-party Bell inequality $BI_+=2$ with local bound
$Q_+=2$, independent of the performed measurements of Bob and
Charlie. On the other hand, in the case of outcome $A_2=-1$, the
resulting two-party Bell inequality is $BI_-=2CHSH_{BC}-2$. Because
$BI_+$ does not depend on the actual form of Bob's
and Charlie's measurements, we can apply the Horodecki formula
\cite{HHH} for the calculation of the CHSH value corresponding to
the state~(\ref{rho2}), which turns out to be
$CHSH_{BC}=2.693620$.
This way, we obtain the quantum value
$Q_-=2CHSH_{BC}-2=3.387240$. 
Hence, the overall maximum is $Q=p_0Q_+ + (1-p_0)Q_- = 2.334184$, 
entailing the ratio $Q/L= 1.167092$. 
Interestingly, Bob's and Charlie's settings now require complex
numbers, whereas in the case of the two-body Bell inequality of
section~\ref{two-body} it was enough to consider real-valued
measurements.

We wish to note that using the software developed in \cite{gruca}
based on the geometric method \cite{kz}, we could not find a
better $Q/L$ ratio with our state $\varrho^*$ up to five
measurement settings per party. Hence, we conjecture that the
ratio $Q/L=1.167092$ found for $\varrho^*$ is optimal or at least
very close to optimality for any number of measurement settings.

\vspace{5 mm}

\section{Conclusion}\label{conc}
We have provided an affirmative answer to the following open
question: Is there an example of a set of separable two-party
marginals, such that any global state compatible with these
marginals is nonlocal, witnessed by violation of a Bell
inequality? We found such a state, which is, in fact, uniquely
determined by its two-party reduced states. Among two-setting Bell
inequalities,
this state is maximally violated by Sliwa's inequality 4 
giving a ratio of $1.167092$ 
for the quantum per classical bound. Interestingly, the same state
also violates a Bell inequality built up from only two-body
correlation terms. An intriguing open question is whether our
result could be strengthened by considering the stronger notion of
the genuine nonlocality scenario \cite{svet} instead of the standard
nonlocality scenario considered in the present study. That is, we
inquire whether there exist two-party separable marginals such
that any global three-party state compatible with these marginals
is genuinely tripartite nonlocal. The state we considered here was
not genuinely multipartite entangled hence cannot lead to genuine
nonlocality. Therefore, new insight is very likely needed to
tackle this interesting open problem.

\section*{Acknowledgments}

We would like to thank R. Augusiak and J. Tura for valuable
discussions. W.L. is supported by Polish Ministry of Science and Higher Education Grant
no.~IdP2011~000361 and the Foundation for Polish Science TEAM project
co-financed by the EU European Regional Development Fund.
T.V. acknowledges financial support from the J\'anos Bolyai Programme of the 
Hungarian Academy of Sciences and from the Hungarian National Research 
Fund OTKA (PD101461). The publication was supported by the 
T\'AMOP-4.2.2.C-11/1/KONV-2012-0001 project. The project has been supported by 
the European Union, co-financed by the European Social Fund.

\end{document}